\documentclass[12pt]{article}
\usepackage{amsthm}
\usepackage{graphicx}
\usepackage{enumerate}
\usepackage{natbib, threeparttable}
\usepackage{url} 
\usepackage{setspace, bm} 

\usepackage{amsmath,mathrsfs}
\usepackage{enumerate}

\usepackage{multirow}
\usepackage{verbatim}
\usepackage[font=footnotesize]{subcaption}
\usepackage{amssymb}
\usepackage{tikz}
\usepackage{booktabs}
\usepackage[normalem]{ulem} 

\usepackage{changepage}

\newcommand{\blind}{1}

\theoremstyle{plain}

\theoremstyle{remark}

\begin{document}

\def\spacingset#1{\renewcommand{\baselinestretch}%
{#1}\small\normalsize} \spacingset{1}


\if1\blind
{
  \title{\bf Bayesian Estimation of Propensity Scores for Integrating Multiple Cohorts with High-Dimensional Covariates}
   \author{Subharup Guha\thanks{This work was supported by 
 the National Science Foundation under award DMS-1854003 to SG, and by 
the National Institutes of Health under award CA269398 to SG and YL, and awards  CA209414 and CA249096  to YL.
}
\hspace{.2cm}\\
    Department of Biostatistics,  University of Florida\\
    and \\
    Yi Li\\
    Department of Biostatistics, University of Michigan}
  \maketitle
} \fi

\bigskip
\begin{abstract}
Comparative meta-analyses of  groups of subjects by integrating multiple observational studies rely on estimated propensity scores (PSs) to mitigate covariate imbalances. However, PS estimation grapples with the theoretical and practical challenges posed by high-dimensional covariates.  Motivated by an integrative analysis of breast cancer patients across seven medical centers, this paper tackles the challenges associated with integrating multiple observational datasets. 
The proposed inferential technique, called \underline{B}ayesian \underline{M}otif \underline{S}ubmatrices for \underline{C}ovariates (B-MSC), addresses the curse of dimensionality by a hybrid of Bayesian and frequentist approaches.  B-MSC uses nonparametric Bayesian ``Chinese restaurant" processes to eliminate redundancy in the high-dimensional covariates and discover latent \textit{motifs} or lower-dimensional structure. With these motifs as potential predictors, standard regression techniques can be utilized to accurately infer the PSs and facilitate covariate-balanced group comparisons. 
Simulations and meta-analysis of the motivating cancer investigation demonstrate the efficacy of the B-MSC approach to accurately estimate the propensity scores and efficiently address covariate imbalance when integrating observational health studies with high-dimensional covariates.
\end{abstract}

\noindent%
\vfill

\newpage
\spacingset{1.45} 

\section{Introduction} \label{S:intro}
 The primary goal of integrative analysis across multiple observational health studies is to compare two or more exposure groups, delineated, for instance, by risk behavior, disease subtype, or treatment. Covariate imbalance can introduce bias into group comparisons \citep{smith2018analysis, robins1995semiparametric, Rosenbaum_Rubin_1983, Li_etal_2018}, making covariate balance essential for valid comparisons \citep{robins1995semiparametric, Rosenbaum_Rubin_1983}. However, achieving covariate balance becomes more challenging when dealing with high-dimensional covariates.

This work is motivated by a  multiple-study  
 breast cancer investigation of female patients  at seven nation-wide medical centers such as Mayo Clinic, 
University of Pittsburgh, and University of Miami.   The  data are publicly
    available from The Cancer Genome Atlas (TCGA) portal \citep{GDC} and include 30      demographic and clinicopathological attributes,  20,531 mRNA expression levels and 24,776 copy number alteration (CNA)  measurements. Some  study-specific  summaries are presented in Table 1 of the Appendix. 
        The  inferential focus is the covariate-balanced comparisons of overall survival between the  disease subtypes infiltrating ductal carcinoma (IDC) and infiltrating lobular carcinoma (ILC), two major subtypes of breast cancer, by integrating the cancer cohorts from these seven centers. 
   This knowledge may inform the development of viable guidelines for regulating targeted therapies and precision medicine among the breast cancer population, specifically tailored to different disease subtypes
\citep{schmidt2016precision}.  
However, the highly unbalanced  groups in Table~1 of the Appendix present a challenge by  confounding naive group comparisons.  Indeed, the literature present conflicting findings due to the disease subtype's confounding with, for example, cancer stage \citep{barroso2016differences}. Therefore, it is imperative to implement covariate-balancing.  The high-dimensional biomarkers, featured by these studies, pose additional challenges when ensuring covariate balance and analyzing data.

When analyzing single observational studies with two groups, 
  the  propensity score  (PS)
 \citep{Rosenbaum_Rubin_1983} is critical for covariate-balancing methods such as weighting and matching, and   can be estimated   in a robust manner \citep{ shu2021estimating}. 
 However, propensity scores are not appropriate for data integration in multigroup, multistudy settings. 
In these investigations, \cite{guha2023causal} achieved covariate-balanced inferences by    generalizing the PS  to  the \textit{observed population multiple propensity score} (o-MPS), defined as the  probability of a patient or subject belonging to a study-group combination given their covariates. 
Due to their pivotal role  in covariate-balanced data integration by weighting or matching,
the unknown o-MPSs must  be estimated. This is achieved by regressing the study-group combination on the covariates using  frequentist or  Bayesian   methods. Ridge regression \citep{hilt1977ridge}, lasso \citep{Tibshirani_1997}, adaptive lasso \citep{Zou_2006}, and group lasso \citep{meier2008group} are arguably the most popular regularization techniques  in the frequentist causal inference paradigm.  Bayesian  methods for  covariate-balanced inference are reviewed by \cite{linero2023and}, \cite{li2023bayesian}, \cite{oganisian2021practical}, and \cite{hill2020bayesian}, and  include  regularization methods   \citep[e.g.,][]{antonelli2019high,BaLASSO,zigler2016central} and 
Bayesian Additive Regression Tree (BART) models  \citep{chipman2010}.  However, these methods are often not  as effective when the number of  covariates is large.

\paragraph{High-dimensional covariates}
While accurate o-MPS estimates are crucial to achieve covariate balance, they encounter significant challenges in high-dimensional observational studies such as the motivating TCGA breast cancer application. The pervasive collinearity in ``small $N$-large $p$" regression settings leads to unstable estimation and erroneous out-of-the-bag predictions.  Generally, in high-dimensional regression settings, the  predictors become virtually unidentifiable without robust priors, making it difficult to identify a sparse subset of covariates  \citep{guha2016nonparametric}.   In TCGA studies, numerical collinearity arises due to the singularity of the sample variance-covariance matrix. This occurs because the number of covariates, such as demographic factors, clinicopathological data, mRNA expression, and CNA measurements, exceeds the sample size.
  Identifiability issues in regression coefficients prevent the determination, based solely on a likelihood function, of the true association of one or both these covariates with o-MPS.

For tackling these critical impediments to the accurate estimation of o-MPS  as a first step of data integration,  we propose  a new Bayesian inferential procedure  called \underline{B}ayesian \underline{M}otif \underline{S}ubmatrices for \underline{C}ovariates (B-MSC). The method leverages nonparametric Bayesian Chinese restaurant processes (CRPs) \citep{Lijoi_Prunster_2010, muller2013bayesian}, and effectively eliminates redundancy in high-dimensional covariates. The method exhibits several advantages in comparison to existing techniques. Departing from existing Bayesian  methods \cite[e.g.,][]{linero2023and, li2023bayesian}, our method  is not ``dogmatically Bayesian,'' but rather  a  hybrid of Bayesian and frequentist solutions that permits flexible, out-of-the-box use of available software for regularization. Specifically, the proposed Bayesian nonparametric methods effectively mitigate the curse of dimensionality  
and unveil the latent structure or \textit{motif} in the covariates. 
Then,  using these identified motif elements as potential predictors, B-MSC applies existing frequentist or Bayesian regression techniques  for lower-dimensional settings to estimate the o-MPS. This allows accurate and computationally efficient estimation of o-MPS  and, subsequently, facilitates integrative comparative analyses of retrospective  cohorts via weighting or matching methods. 
 
Section \ref{S:stage0} introduces some basic notation and theoretical assumptions.   
 Section \ref{S:B-MSC}  describes the B-MSC hierarchical model and prior. A Bayesian inferential procedure, including a fast-mixing MCMC algorithm, is described in Section~\ref{S:inference}. Section \ref{S:survival} applies the proposed techniques to make integrative group comparisons with right-censored  outcomes and high-dimensional covariates. We conduct simulations in 
 Section \ref{S:simulation} to demonstrate the efficacy of the  B-MSC approach in dimension reduction and   o-MPS estimation. Section \ref{S:data analyses} analyzes the high-dimensional  
 TCGA breast cancer studies.  
Section \ref{S:discussion} concludes with some remarks.



\section{Bayesian Propensity Score Estimation for Meta-Analysis of Retrospective  Cohorts with High-Dimensional Covariates}\label{S:stage0}

The investigation  comprises $J$ observational studies and focuses on comparing $K$ groups.  We  assume   $J$ and $K$ are  small; in the context of the TCGA  database discussed earlier,  $J=7$ and $K=2$.  For subject $i=1,\ldots,N$, let $Z_i  \in \{1,\ldots,K\}$ denote
   their groups and  $S_i  \in \{1,\ldots,J\}$  be the 
  observational study to which the $i$th subject belongs.  
The database contains a large number of  $p$    covariates shared by all $J$ studies.  For the $i$th subject,  let the vector of covariates, $\mathbf{x}_i$, belong to a space $\mathcal{X} \subset \mathcal{R}^p$ shared by  the $JK$ groups and studies. Further, let $\mathbf{x}_i=(\mathbf{x}_{i}^{[1]},\mathbf{x}_{i}^{[2]})$, 
where   vector $\mathbf{x}_{i}^{[1]}$ consists of    $p_1$ continuous    covariates, and  $\mathbf{x}_{i}^{[2]}$ contains   $p_2$ factor-valued  covariates belonging to the set $\{1,2,\ldots,A\}$ for an integer $A>1$, which can be ordinal or categorical; thus, $p=p_1+p_2$. \cite{guha2023causal}  defines the {observed population multiple propensity score} or o-MPS    as  
$\delta_{sz}(\mathbf{x})    =  
\bigl[S=s,Z=z \mid \mathbf{X} = \mathbf{x}\bigr]_{+}
$
for   $(s,z) \in \Sigma$
$\equiv$ $\{1,\ldots,J\} \times \{1,\ldots,K\}$
and  $\mathbf{x} \in \mathcal{X}$.  
 In many applications, both $p_1$ and  $p_2$ are   large, and $\min\{p_1, p_2\}$  far exceeds $N$. Stacking these $N$ row vectors, we obtain an $N\times p_1$ matrix, $\mathbf{X}^{[1]}$, of continuous covariates and an $N \times p_2$ matrix, $\mathbf{X}^{[2]}$, of factor covariates. Since the covariates  are subsequently used in regression settings, the columns of continuous submatrix $\mathbf{X}^{[1]}$ are often empirically standardized to zero means and unit standard deviations.
 It is trivial to extend this framework to accommodate additional factor variables, e.g., with binary and trinary covariates,   we set $\mathbf{x}_i=(\mathbf{x}_{i}^{[1]},\mathbf{x}_{i}^{[2]},\mathbf{x}_{i}^{[3]})$ with $A_2=2$ and $A_3=3$. Denote by $T_i^{(z)}$  the     counterfactual
 outcome  if $Z_i$ were $z$.  The realized outcome is $T_i= T_i^{(Z_i)}$. Denoting by  $C_i$ the censoring time, $Y_i= \min\{T_i^{(Z_i)},C_i\}$ is the observed survival time with event indicator $\vartheta_i=\mathcal{I}(T_i^{(Z_i)}\le C_i)$, where $\mathcal{I}(\cdot)$ is an indicator function. 
 In the TCGA  datasets, submatrices $\mathbf{X}^{[1]}$ and $\mathbf{X}^{[2]}$ of the covariate matrix $\mathbf{X}$ include high-dimensional demographic and clinicopathological variables in addition to mRNA and CNA biomarkers. The  CNA measurements  are coarsened as binary factors: 1 (no CNA) or 2 (some CNA), i.e., $A=2$.  Also in this context,  $T^{(1)}_i$ and $T^{(2)}_i$   represent  the counterfactual outcomes if patient $i$ were disease subtypes IDC and ILC, respectively; $\vartheta_i=1$ if $Y_i$,
  the observed survival time of the $i$th patient,
  is uncensored, and equals $0$ otherwise.

As   index ($i$) contains no meaningful information,
the individual measurements are a random sample from an   \textit{observed   distribution},  $[S,Z,\mathbf{X},T]_{+}$, where the symbol $[\cdot]_{+}$  represents distributions or densities under the observed population.   
 Following  \cite{Imbens_2000},
we make the following assumptions: (a)~\textbf{Stable unit treatment value}:  a  subject’s study and group memberships do not influence the potential outcomes of any other subject. Furthermore, each subject has $K$ potential outcomes of which only one is observed; (b) \textbf{Study-specific  weak unconfoundedness}: Conditional on  study $S$ and covariate  $\mathbf{X}$,  the event  $[Z=z]$  is independent of   counterfactual   outcome $T^{(z)}$ for all $z=1,\ldots,K$; 
 and (c) \textbf{Positivity}: Joint density  $[S=s,Z=z,\mathbf{X}=\mathbf{x}]_{+}$ is   strictly positive for all $(s,z,\mathbf{x})$, ensuring  that the study-group memberships and   covariates do not have non-stochastic, mathematical relationships.
 
\subsection{Bayesian motif submatrices for dimension reduction}\label{S:B-MSC}

 B-MSC utilizes the sparsity-inducing property of Bayesian mixture models to detect lower-dimensional  structure in the covariate submatrices. We  perform   bidirectional, unsupervised global clustering of $\mathbf{X}^{[1]}$ and $\mathbf{X}^{[2]}$. 
A nonparametric Chinese restaurant process (CRP), denoted by $\mathscr{C}^{[1]}_r$,   discovers $q_r^{[1]}$ latent  subpopulations or \textit{cliques} among the $N$ subjects (submatrix rows). Simultaneously,  another   CRP, $\mathscr{C}^{[1]}_c$, discovers  $q_c^{[1]}$ latent  \textit{clusters} among the $p_1$ continuous biomarkers (submatrix columns). 
This gives a  ``denoised'' lower-dimensional version of submatrix $\mathbf{X}^{[1]}$ called the   \textit{motif submatrix},  $\mathbf{\Phi}^{[1]}$, of dimension $q_r^{[1]} \times q_c^{[1]}$.  
For the factor covariate matrix  $\mathbf{X}^{[2]}$, another set of CRPs, $\mathscr{C}^{[2]}_r$ and $\mathscr{C}^{[2]}_c$, performs global unsupervised clustering of the rows and columns of $\mathbf{X}^{[2]}$  to give a $q_r^{[2]} \times q_c^{[2]}$ motif submatrix,  $\mathbf{\Phi}^{[2]}$. 

CRPs attempt to reverse the curse of dimensionality by guaranteeing with high  probability that $q_r^{[1]} \ll N$ and $q_c^{[1]} \ll p_1$, while   allowing  the number of cliques and clusters to be  apriori unknown.
Instead of the highly collinear covariates  $\mathbf{X}=(\mathbf{X}^{[1]},\mathbf{X}^{[2]})$ as   predictors of the study-group memberships, estimates of the    compact motif matrices   play the role of covariates for reliably inferring the o-MPS, as demonstrated later by  simulation studies. The meta-analytical weighting methods outlined in Introduction are then  employed to make covariate-balanced comparisons of the health outcomes of the $K$  groups.

For $t=1,2$, biomarker clusters are likely to exist in submatrix $\boldsymbol{X}^{[t]}$ because  
$N$ is much smaller than $p_t$; since the rank of $\boldsymbol{X}^{[t]}$ is $N$ or less,  its $p_t$  columns have  high redundancy. Closely related to some aspects of the proposed B-MSC approach are  global  clustering algorithms  for continuous covariate matrices \citep[e.g.,][]{Dahl_2006,Muller_Quintana_Rosner_2011}, which assume that the matrix rows and columns   can be  independently  permuted to reveal the  underlying lower-dimensional signal.
Alternatively, B-MSC can  be viewed as an adaptation of  local clustering algorithms for continuous covariates \citep{guha2016nonparametric,Lee_etal_2013} and binary covariates \citep{guha2022predicting} to multiple factor levels and computationally  efficient global clustering implementations. From a scientific perspective, biomarkers do not act in isolation but in concert to perform  biological functions, resulting in similar biomarker profiles \citep[e.g.,][]{mundade2014genetic}. Biomarkers exhibiting   similar patterns in high-dimensional genomic, epigenomic, and microbiome data   have been  exploited  to achieve dimension reduction via  mixture models \citep[e.g.,][]{Kim_etal_2006, Dunson_etal_2008}.  
 However, to our knowledge,  this  phenomenon has not  been fully utilized to achieve  efficient  inferences in covariate-balanced integrative analyses.

More formally, for the columns of covariate submatrix $\boldsymbol{X}^{[t]}$, where $t=1,2$, we envision \textit{biomarker-cluster  mapping variables}, $c_1^{[t]},\ldots,c_{p_t}^{[t]}$, with  $c_i^{[t]}=u$ representing the event that the $j${th} biomarker of $\boldsymbol{X}^{[t]}$ is allocated to the $u${th} latent cluster, for biomarker $j=1,\ldots,p_t$, and  cluster  $u=1,\ldots,q_c^{[t]}$. For a positive mass parameter $\alpha_{c}^{[t]}$, CRP prior $\mathscr{C}^{[t]}_c\bigl(\alpha_{c}^{[t]}\bigr)$  assigns the following PMF to 
 the vector of $p_t$ cluster mapping variables:
\[
    \bigl[\boldsymbol{c}^{[t]} \mid \alpha_{c}^{[t]}\bigr] = \frac{\Gamma(\alpha_{c}^{[t]}) \bigl(\alpha_{c}^{[t]}\bigr)^{q_c^{[t]}}}{\Gamma(\alpha_{c}^{[t]} + p_t)} \prod_{u=1}^{q_c^{[t]}} \Gamma(m_u^{[t]}), \quad \boldsymbol{c}^{[t]} \in \mathscr{Q}_{p_t}, \label{eq:CRP}
    \]
 where $m_u^{[t]}$ is the  number of biomarkers belonging to the $u$th latent cluster in  mapping vector $\boldsymbol{c}^{[t]}$, and  $\mathscr{Q}_{p}$  is the set of all possible partitions of  $p$ objects into one or more sets \citep{muller2013bayesian}. 
 Dimension reduction occurs because the random number of  clusters, $q_c^{[t]}$, is approximately equal to $\alpha_{c}^{[t]} \log(p_t)$ as $p_t$ is sufficiently large \citep{Lijoi_Prunster_2010}. 
  
Analogously, for the $N$ rows of covariate submatrix $\boldsymbol{X}^{[t]}$,  there exists \textit{subject-clique  mapping variables}, $r_1^{[t]},\ldots,r_n^{[t]}$, with  $r_i^{[t]}=u$ representing the event that the $i${th} subject is allocated to the $u${th} latent clique, where $i=1,\ldots,N$, and    $u=1,\ldots,q_r^{[t]}$.  For a positive mass parameter $\alpha_{r}^{[t]}$, the vector of clique mapping variables is given a CRF prior, $\boldsymbol{r}^{[t]}\sim \mathscr{C}^{[t]}_r\bigl(\alpha_{r}^{[t]}\bigr)$, for which the unknown number of cliques, $q_r^{[t]}$, is  asymptotically lower order than $N$. 

\paragraph{Motif submatrices}  Unlike the CRP allocation models for dimension reduction, the priors for  motif submatrices $\mathbf{\Phi}^{[1]}$ and $\mathbf{\Phi}^{[2]}$ depend on the covariate types (continuous or factor) of submatrices $\boldsymbol{X}^{[1]}$ and $\boldsymbol{X}^{[2]}$. Consider motif submatrix  $\mathbf{\Phi}^{[1]}=(\phi_{rc}^{[1]})$. 
We assume $\phi_{rc}^{[1]} \stackrel{\text{i.i.d.}}\sim N(0,\tau^2)$  for $r=1,\ldots,q_r^{[1]}$ and   $c=1,\ldots,q_c^{[1]}$, with  $\tau^2$ assigned an inverse gamma hyperprior. 
Next, consider motif submatrix  $\mathbf{\Phi}^{[2]}=(\phi_{rc}^{[2]})$. Submatrix   $\boldsymbol{X}^{[2]}$ consists of  factors with levels belonging to $\{1,2,\ldots,A\}$. Therefore, for $r=1,\ldots,q_r^{[2]}$ and   $c=1,\ldots,q_c^{[2]}$, we assume $\phi_{rc}^{[2]} \stackrel{\text{i.i.d.}}\sim \mathscr{B}_A(\boldsymbol{g})$, a generalized Bernoulli or categorical distribution on $A$ categories, with  the probability vector $\boldsymbol{g}=(g_1,\ldots,g_A)$  given a  Dirichlet distribution hyperprior. 
This implies that   motif submatrix  $\mathbf{\Phi}^{[2]}$ also consists of factors taking values in $\{1,2,\ldots,A\}$.   

\paragraph{Likelihood functions for covariates}  We  model the   matrix $\boldsymbol{X}^{[1]}$ elements as noisy versions of the mapped elements of $\mathbf{\Phi}^{[1]}$.   We specify distributional assumptions  guaranteeing that   biomarkers in a cluster  have similar column profiles in $\boldsymbol{X}^{[t]}$ and  subjects in a clique have similar row profiles in $\boldsymbol{X}^{[t]}$. The elements of  submatrix $\boldsymbol{X}^{[1]}$  are  conditionally Gaussian: 
$x_{ij}^{[1]} \mid \mathbf{\Phi}^{[1]}, \boldsymbol{c}^{[1]}, \boldsymbol{r}^{[1]} \stackrel{\text{indep}}\sim N(\phi_{r_i^{[1]} c_j^{[1]}}^{[1]}, \sigma^2)$, where $i=1,\ldots,N,$ $j=1,\ldots,p_1$, and 
 $\sigma^2$ has a truncated inverse gamma hyperprior that   ensures  $R^2$ is sufficiently large. At first glance, the Gaussian   likelihood   and  common  variance  $\sigma^2$ may appear to be a strong parametric assumption. However, if the biomarker and subject labels are non-informative, the assumed CRP priors for the clusters and cliques constitute a semiparametric model for the $\boldsymbol{X}^{[1]}$ elements and the arbitrary, true underlying distribution of the i.i.d.\ submatrix   elements is  consistently inferred a posteriori \citep{Ghosal_Ghosh_Ramamoorthi_1999}.

The matrix  $\boldsymbol{X}^{[2]}$ elements as possibly corrupted versions of the mapped motif elements  with  low probabilities of \textit{corruption} or   covariate-motif mismatch, i.e.,  $x_{ij}^{[2]} \neq \phi^{[2]}_{r_i^{[2]}\,c_j^{[2]}}$.   
The elements of factor submatrix $\boldsymbol{X}^{[2]}$ are related to the mapped elements of  $\mathbf{\Phi}^{[2]}$ as follows:
$P\left(x_{ij}^{[2]}=x \mid \phi_{r_i^{[2]} c_j^{[2]}}^{[2]}=\phi, \mathbf{\Phi}^{[2]}, \boldsymbol{c}^{[2]}, \boldsymbol{r}^{[2]}, \boldsymbol{W} \right)  =
    w_{\phi x}$, 
    $\phi,x \in \{1,2,\ldots,A\}$
for a \textit{corruption probability matrix} $\boldsymbol{W}=(w_{\phi x})$ of dimension  $A \times A$.
Low  corruption levels are achieved by a diagonally dominant  $\boldsymbol{W}$. 
Since matrix $\boldsymbol{W}$ is row-stochastic,  row vectors $\boldsymbol{w}_1,\ldots,\boldsymbol{w}_A$ are assigned independent priors on the unit simplex in $\mathcal{R}^A$. Let  $\boldsymbol{1}$ be the vector of $A$ ones. For $\phi=1,\ldots,A$, let  $\boldsymbol{1}_\phi$ be the  vector in $\mathcal{R}^A$  with  the  $\phi$th element equal to 1 and the other $(A-1)$ elements equal to zero.  The  $\phi$th row vector of $\boldsymbol{W}$ is
$ \boldsymbol{w}_\phi = l_\phi \boldsymbol{1}_\phi + (1-l_\phi) \tilde{\boldsymbol{w}}_\phi$, where
  $\tilde{\boldsymbol{w}}_\phi \sim \mathscr{D}_A\left(\alpha\boldsymbol{1}/A\right)$ and   
   $l_\phi \sim \text{beta}(l_\alpha, l_\beta)\cdot\mathcal{I}(l_{s}>l^*)$, 
for prespecified constants $l^*$, $l_\alpha$ and $l_\beta$, and  $\mathscr{D}_A$ representing a Dirichlet distribution in~$\mathcal{R}^A$. The condition $l^*>0.5$ implies $\boldsymbol{W}$ is diagonally dominant. Through extensive simulations, we find that  $l^*>0.85$  produces sufficiently ``tight'' clusters and cliques in  submatrix  $\boldsymbol{X}^{[2]}$.

\subsection{Estimating o-MPS in the presence of high-dimensional covariates}\label{S:inference}

For integrative covariate-balanced inferences, we propose a hybrid Bayesian and frequentist approach for o-MPS estimation.  
Since the o-MPS is $\delta_{sz}(\mathbf{x})    =  
\bigl[S=s,Z=z \mid \mathbf{X} = \mathbf{x}\bigr]_{+}$    
 we could estimate it in theory by regressing  $(S_i,Z_i)$ on  the $p$-dimensional covariate $\mathbf{x}_i$  ($i=1,\ldots,N$). Indeed, for low-dimensional covariates, 
the o-MPS is accurately estimated  using  parametric (e.g., multinomial logistic)  or nonparametric, Bayesian or frequentist regression models.  Unsurprisingly, 
this strategy is  inefficient or even untenable in the presence of high-dimensional covariates.
 When $p$ is large,  we propose the following two-step inferential procedure using the lower-dimensional approximation of   $\mathbf{x}_i$ as  predictor: 

 \smallskip

\noindent \textbf{Step 1} \quad We first obtain MCMC estimates $\hat{\mathbf{\Phi}}^{[1]}$ and $\hat{\mathbf{\Phi}}^{[2]}$ of the lower-dimensional motif submatrices   and the least squares allocations $\hat{\boldsymbol{r}}^{[1]}$ and $\hat{\boldsymbol{r}}^{[2]}$ of the clique mapping variables:

\smallskip

\textit{Step 1a} \quad 
Following initialization using  naive estimation strategies, the B-MSC model parameters are iteratively updated using MCMC techniques until the chain converges to the posterior.  
 In the Appendix, we summarize a computationally efficient, fast-mixing  MCMC algorithm. Exploiting the B-MSC model structure, the MCMC sampler can be parallelized to separately update the non-intersecting parameters  related to submatrices~$\mathbf{X}^{[1]}$ and $\mathbf{X}^{[2]}$. 
 Using the post-burn-in  MCMC samples, Bayes estimates  are  computed for the posterior probability of  each biomarker pair belonging to the same cluster and each subject pair belonging to the same clique. Following \cite{Dahl_2006}, these probabilities are  used to compute point estimates for the cluster and clique mapping variables, called  \textit{least-squares  allocations}. The estimated cluster variables are denoted by $\hat{\boldsymbol{c}}^{[1]}$ and $\hat{\boldsymbol{c}}^{[2]}$. The estimated clique variables are denoted by $\hat{\boldsymbol{r}}^{[1]}$ and $\hat{\boldsymbol{r}}^{[2]}$.

\textit{Step 1b} \quad Setting   the cluster and clique mapping variables equal to their   least-squares  allocations, a second MCMC sample is generated and post-processed 
     to obtain  Bayes estimates of  motif submatrices, denoted by $\hat{\mathbf{\Phi}}^{[1]}$ and~$\hat{\mathbf{\Phi}}^{[2]}$.

 \smallskip

\noindent \textbf{Step 2} \quad 
Conditional on  Bayes estimates $\hat{\mathbf{\Phi}}^{[1]}$ and $\hat{\mathbf{\Phi}}^{[2]}$ of the lower-dimensional motif submatrices   and the least squares allocations $\hat{\boldsymbol{r}}^{[1]}$ and $\hat{\boldsymbol{r}}^{[2]}$ of the clique mapping variables, the B-MSC approach  regresses $(S_i,Z_i)$ on the subject-specific motif vector   $\bigl(\hat{\boldsymbol{\phi}}^{[1]}_{\hat{\boldsymbol{r}}_i^{[1]}}, \hat{\boldsymbol{\phi}}^{[2]}_{\hat{\boldsymbol{r}}_i^{[2]}}\bigr)$ of length $q_c^{[1]} + q_c^{[2]} \ll p$. 
Commonly used regression techniques for low- or high-dimensional covariate-balanced inference, e.g., multinomial logistic regression,  random forests, and BART,   are then more   efficiently employed to  discover the relationship between  the study-group memberships and the clique-specific rows of the motif matrices.  In this manner, we obtain
o-MPS estimate $\hat{\delta}_{sz}(\mathbf{x}_i)=\hat{\delta}_{sz}\bigl(\hat{\boldsymbol{\phi}}^{[1]}_{\hat{\boldsymbol{r}}_i^{[1]}}, \hat{\boldsymbol{\phi}}^{[2]}_{\hat{\boldsymbol{r}}_i^{[2]}}\bigr)$, for $s=1,\ldots,J$, $z=1,\ldots,K$.


\section{Application: Covariate-balanced survival function  estimation by meta-analyzing right-censored outcomes}\label{S:survival}

After the o-MPS estimates are available, weighting methods may be applied to achieve covariate-balanced inferences. In single observational studies, since weighting inferences utilizing inverse probability  weights (IPWs) may be   unstable when some subjects have very small PSs \citep{li2019propensity}, 
several authors \citep[e.g.,][]{crump2006moving,li2013weighting,Li_etal_2018,li2019propensity}  have  proposed   alternative PS-based weighting strategies.    
\cite{guha2023causal} introduced a general  framework that extends several 
weighting methods in the literature to effectuate covariate-balanced integrative analyses of multiple   cohorts with multiple groups. We briefly summarize the  methodology here. 
For any   weighting method, the \textit{unnormalized weight function}, denoted by    $\tilde{\rho}(s,z,\mathbf{x})$, and  the  \textit{empirically  normalized   balancing weight} of the $i$th subject,  $ \bar{\rho}_i= N \tilde{\rho}(s_i,z_i,\mathbf{x}_i)/\sum_{l=1}^N \tilde{\rho}(s_l,z_l,\mathbf{x}_l)$, can be computed, producing sample weights that sum to $N$. For example,  $\tilde{\rho}(s,z,\mathbf{x})=1/\hat{\delta}_{sz}(\mathbf{x})$ extends IPWs   and generalized IPWs \citep{Imbens_2000} to    \textit{integrative combined} (IC) weights
 in meta-analytical settings. Similarly,    generalized overlap weights \citep{li2019propensity} can be extended  to    \textit{integrative generalized overlap} (IGO) weights by assuming $\tilde{\rho}(s,z,\mathbf{x})=$ $\hat{\delta}_{sz}^{-1}(\mathbf{x})/\sum_{s'=1}^J\sum_{z'=1}^K\hat{\delta}_{s'z'}^{-1}(\mathbf{x})$. In addition to incorporating equal amounts of information from the $J$ studies, most weighting methods provide accurate inferences only for hypothetical  pseudo-populations in which the $K$ groups are equally prevalent. 
 For natural populations with unequally distributed groups, \cite{guha2023causal} developed  FLEXOR weights to create more realistic  pseudo-populations.
Suppose the relative group prevalence in the natural population of interest  is the probability vector, $\boldsymbol{\theta}=(\theta_{1},\ldots,\theta_{K})$, e.g., in the  TCGA breast cancer studies,    $\boldsymbol{\theta}=$ $(8/9,  1/9)$ matches the known U.S. proportions  of  breast cancer subtypes IDC and ILC.  Then  $\tilde{\rho}(s,z,\mathbf{x})=$ $\hat{\delta}_{sz}^{-1}(\mathbf{x})\bigl( \sum_{s'=1}^J  \sum_{z'=1}^K \frac{\breve{\gamma}_{s'} ^2\theta_{z'}^2}{\hat{\delta}_{s'z'}(\mathbf{x})}  \bigr)^{-1}$ gives the FLEXOR pseudo-population with the  characteristics: \textit{(i)} the relative weights of the groups matches that of the natural population; \textit{(ii)} probability vector $\breve{\boldsymbol{\gamma}}=(\breve{\gamma}_{1},\ldots,\breve{\gamma}_{J})$ represents   optimal study weights and is easily estimated by an efficient numerical procedure with negligible computational costs; and \textit{(iii)} FLEXOR maximizes the effective sample size in a broad-ranging family  encompassing many meta-analytical weighting methods, including  IC and IGO weights.

\paragraph{Covariate-balanced weighted survival analysis} \quad
As an application, 
we  apply  the proposed o-MPS estimation, in conjunction with
the weighting methods of \cite{guha2023causal}, to analyze right-censored outcomes with unbalanced high-dimensional covariates.  Consider   survival functions of the counterfactual outcomes  $T^{(z)}$  in the pseudo-population:
 $S^{(z)}(t) = {\text{P}}\bigl[T^{(z)} > t\bigr]$  for $t>0$ and  group $z=1,\ldots, K$. 
As previously noted,   realized outcome $T= T^{(Z)}$.
Analogously to the  weak unconfoundedness assumption for the observed population, we   
make an identical  assumption for the pseudo-population.  That is,
 $[T | S, Z, \mathbf{X}] =$ 
    $[T | S, Z, \mathbf{X}]_+$, where 
  $[\cdot]$ denotes pseudo-population densities.
Unlike the observed population, the covariate-balanced  pseudo-population
gives us $
    [T \mid Z=z] = [T^{(z)}]$,
 facilitating   weighted estimators of different features of  pseudo-population potential outcomes. 
 Specifically,
  using the empirically normalized generalized balancing weights,  the single-study estimator of \cite{xie2005adjusted} can be extended in a straightforward manner to  obtain  the \textit{balance-weighted  Kaplan-Meier estimator} (BKME) of  pseudo-population survival function $S^{(z)}(t)$ as follows.  
Among the $N$ subjects, suppose the observed failures, with possible ties, occur 
at the distinct times $0<t_1<\ldots<t_D$. For the $z$th group,  using the empirically normalized generalized balancing weights, the weighted number of deaths and the weighted number of subjects at risk at time $t_j$  are, respectively,
$d_j^{(z)}=N\sum_{i:Y_i=t_j, \vartheta_i=1}  \bar{\rho}_i\, \mathcal{I}(Z_i=z)$  and  $R_j^{(z)}=N\sum_{i:Y_i\ge t_j} \bar{\rho}_i\, \mathcal{I}(Z_i=z),$
 for $j=1,\ldots,D$. 
  The BKME of  the $z$th  survival function in the pseudo-population is then
\begin{equation}
   \hat{S}^{(z)}(t) = 
   \prod_{j:t_j \le t} \bigl(1-d_j^{(z)}/R_j^{(z)}\bigr) 
   \label{BKME}
\end{equation} 
Variance estimate $\widehat{\text{Var}}\bigl(\hat{S}^{(z)}(t)\bigr) =$ $\bigl(\hat{S}^{(z)}(t)\bigr)^2 \prod_{j:t_j \le t} \frac{d_j^{(z)}}{R_j^{(z)}\bigl(R_j^{(z)}-d_j^{(z)}\bigr)}$ is used for  pointwise confidence intervals. 
The BKME  is consistent and asymptotic normal as an estimator of $S^{(z)}(t)$ and  its variance estimate is consistent \citep{fleming2011counting}. 
If some groups are undersampled,   large-sample inferences may not be valid for those groups. Using  $B$ bootstrap samples of size $N$ each,
we could  apply nonparametric bootstrap methods  
to estimate the  standard error  of BKME and construct asymptotic or distribution-free confidence intervals for the group-specific pseudo-population survival functions.



\section{Simulation Study}\label{S:simulation}

We  aimed to evaluate  the effectiveness of the proposed B-MSC approach in \textit{(i)} achieving dimension reduction in $\mathbf{X}$, \textit{(ii)}   inferring the  clique memberships of additional test cases, 
and \textit{(iii)} when used in conjunction with  existing Bayesian or frequentist regression approaches,  inferring the o-MPS of the training and test set subjects, compared to using the same regression approaches  with $\mathbf{X}$ as the high-dimensional predictor. Since the primary focus is ``outcome-free'' dimension reduction and o-MPS estimation, no responses were generated in this simulation study.  
Using the actual $p_1=757$ continuous (750 most variable mRNA biomarkers + 7  clinicopathological) and $p_2=522$ binary (500 most variable CNA biomarkers + 22 binary clinicopathological or socioeconomic) covariates of the seven   TCGA breast cancer studies,  we generated the study-group memberships of 
 $R=500$ simulated datasets. We examined two  scenarios (labeled ``high'' and ``low'') characterized by the degree of association between  o-MPS and its predictors, and determined by a simulation parameter, $\mu$. Each dataset consisted of   $J=4$ observational studies and $K=2$  groups to match the TCGA application.

Specifically, we sampled with replacement  the $p=p_1+p_2=1,279$ covariates  of the TCGA breast cancer patients and randomly allocated the $\tilde{N}=450$ subjects of each artificial dataset  to  $JK=8$ study-group combinations. For  dataset $r=1,\ldots,500$, and  association parameter $\mu$ belonging to $\{10,15\}$, we generated:

 
\begin{enumerate}

\item \textbf{Covariate matrix} \quad  
For  $i=1,\ldots,\tilde{N}$,    covariate vector $\mathbf{x}_{ir}=(x_{i1r},\ldots,x_{ipr})' $ was  sampled with replacement  from the TCGA datasets to obtain  matrix $\mathbf{X}_r$ of dimension $\tilde{N} \times p$, comprising $p=1,280$ binary or continuous covariates.

 \item \textbf{True o-MPS predictors} \quad Without exception, the rank of $\mathbf{X}_r$ was $\tilde{N}$ in all the generated datasets. Let the singular value decomposition of $\mathbf{X}_r$ be $\mathbf{A}_r\mathbf{D}_r{\mathbf{B}_r}'$ where $\mathbf{A}_r$ is an $\tilde{N} \times \tilde{N}$  matrix with orthonormal
columns, $\mathbf{D}_r$ is a $ \tilde{N}\times \tilde{N}$  diagonal matrix with positive diagonal entries, and $\mathbf{B}_r$ is a $p\times \tilde{N}$ 
 matrix with orthonormal columns. 
 Setting $\tilde{n}=50< \tilde{N}$, we let $\hat{\mathbf{X}}_r=$ $\hat{\mathbf{A}}_r\hat{\mathbf{D}}_r(\hat{\mathbf{B}}_r)^T$, where $\hat{\mathbf{D}}_r$  is a $\tilde{n} \times \tilde{n}$ diagonal matrix containing the highest $\tilde{n}$ diagonal elements of $\mathbf{D}_r$, and $\hat{\mathbf{A}}_r$ and $\hat{\mathbf{B}}_r$ extract the matching $\tilde{n}$ columns  of $\mathbf{A}_r$ and $\mathbf{B}_r$. 
 
 Each of the $p$ covariates in matrix $\hat{\mathbf{X}}_r$  was randomly designated as either a  non-predictor,  linear, or quadratic  predictor of the true o-MPS.  For covariate $j=1,\ldots,p$, we  independently generated   category $A_{jr}\in\{0,1,2\}$  with probabilities $0.5$, $0.25$, and $0.25$, respectively, signifying that    the $j$th covariate 
               is a non-predictor, linear  and quadratic predictor. Let  the $\tilde{A}_r=\sum_{j=1}^p \mathcal{I}(A_{jr}>0)$ linear or quadratic  regression predictors derived from $\hat{\mathbf{X}}_r$  be arranged in an $\tilde{N} \times (\tilde{A}_r+1)$ matrix denoted by $\mathbf{Q}_r=(q_{ilr})$, with column 1 consisting of $\tilde{N}$ ones.

\item\label{sz combo} \textbf{True o-MPS and study-group memberships} \quad  Study $s_{ir}$ and group $z_{ir}$ were generated:

          \begin{enumerate}
                \item  \textit{Regression coefficients} \quad Let $(s,z)=(1,1)$ be the reference study-group combination. For the remaining $(JK-1)$  combinations,    regression  vectors 
     $\boldsymbol{\upsilon}_{szr}=(\upsilon_{sz0r},\ldots,\upsilon_{sz\tilde{A}_rr})'
     \stackrel{\text{i.i.d.}}\sim N_{\tilde{A}_r+1}\bigl(\boldsymbol{0}, \boldsymbol{\Sigma} + \mu^2\mathbf{I}_{\tilde{A}_r+1}\bigr)$,
     where $\boldsymbol{\Sigma}^{-1}={\mathbf{Q}_r}'\mathbf{Q}_r$, and let
              $\eta_{iszr} = \sum_{l=0}^{\tilde{A}_r}\upsilon_{szlr}\,q_{ilr}$.
              For the reference combination,  $\eta_{i11r}=0$. Setting $\mu=10$ ($15$) produced   low (high) signal-to-noise ratios.

          \item \textit{True o-MPS}  \quad For the $i$th subject, set    
          $\delta_{szr}(\mathbf{x}_{ir})= \exp(\eta_{iszr})/\sum_{s'=1}^J\sum_{z'=1}^K \exp(\eta_{is'z'r}).$ Evaluate  true o-MPS vector, $\boldsymbol{\delta}_r(\mathbf{x}_{ir})=$ $\{\delta_{szr}(\mathbf{x}_{ir}):s=1,\ldots,J, z=1\ldots,K\}$. 
               
               \item \textit{Study-group memberships} \quad Independently generate   $(s_{ir},z_{ir})$ from the categorical distribution with probability vector $\boldsymbol{\delta}_r(\mathbf{x}_{ir})$.
            
\end{enumerate}

\end{enumerate}


Disregarding knowledge of all  simulation parameters, we  meta-analyzed the   $\tilde{N}$ study-group memberships and $\tilde{N} \times p$ covariate matrix $\mathbf{X}_r$ of each artificial dataset   using the proposed B-MSC methodology. First, the $\tilde{N}=450$ subjects of each dataset were randomly split into training and test samples in a $4:1$ ratio, so that $\tilde{N}_{\text{train}}=360$ and $\tilde{N}_{\text{test}}=90$. The MCMC algorithm of Supplemantary Material was applied to generate posterior samples. We discarded a burn-in  of 10,000 MCMC samples and used  50,000 post-burn-in draws for posterior inferences. Convergence was informally assessed by  trace plots of hyperparameters to determine the appropriate MCMC sample sizes.  The post-burn-in  MCMC draws was processed to estimate  the lower-dimensional motif submatrices $\mathbf{\Phi}^{[1]}$ and $\mathbf{\Phi}^{[2]}$ as described previously. Then, 
the estimated motif submatrices  and  study-group memberships of the $\tilde{N}_{\text{train}}$ subjects were used  to  infer the o-MPS using  ridge regression, lasso, adaptive lasso, group lasso,   random forests, multinomial logistic regression, and BART techniques implemented in  the R packages \texttt{glmnet},\texttt{randomForest}, \texttt{nnet}, and \texttt{BART}, respectively. Posterior estimates of the  latent cliques $\hat{r}_i^{[1]}$ and $\hat{r}_i^{[2]}$ and motif vectors $\bigl(\hat{\boldsymbol{\phi}}^{[1]}_{\hat{\boldsymbol{r}}_i^{[1]}}, \hat{\boldsymbol{\phi}}^{[2]}_{\hat{\boldsymbol{r}}_i^{[2]}}\bigr)$ of length $\hat{q}_r^{[1]}+\hat{q}_r^{[2]}$ of the 
 test set subjects were estimated using the the Appendix MCMC algorithm. Finally, the estimated o-MPS vector $\hat{\boldsymbol{\delta}}(\mathbf{x}_i)$ of length $JK$ for the 
 $\tilde{N}_{\text{test}}$ subjects were computed using only the motif vectors $\bigl(\hat{\boldsymbol{\phi}}^{[1]}_{\hat{r}_i^{[1]}}, \hat{\boldsymbol{\phi}}^{[2]}_{\hat{r}_i^{[2]}}\bigr)$ estimated from their $p$  fully observed covariates, $\mathbf{x}_i$.

For the $p_1=757$ continuous covariates, the average number of estimated  clusters, $\hat{q}_c^{[1]}$,   of the 500 datasets was 173.8 with a standard error of 0.2. For the $p_2=522$ binary covariates, the average number of estimated  clusters, $\hat{q}_c^{[2]}$,   was 38.3 with a standard error of 0.2. In other words, the estimated motif submatrices of B-MSC  were substantially smaller than the full set of covariates. 
Next, for each  dataset, and using the estimated motif matrices of the training samples, we evaluated the  accuracy of the estimated cliques $\hat{r}_i^{[t]}$ of the 
 test samples  using a  measure called the \textit{parity},   $\Delta^{[t]}$, for covariate types $t=1,2$. Since the ``true''  cliques of the TCGA datasets are unknown, the parity compares two quantities: \textit{(a)}  the   estimated clique motifs of the test subjects when the B-MSC model is fitted to  the $\tilde{N}_{\text{train}}$ subjects and applied to the $\tilde{N}_{\text{test}}$ subjects, versus \textit{(b)} the estimated  clique motifs for the same test  subjects  when the B-MSC model is fitted to  all $\tilde{N}$ subjects. A high parity indicates that out-of-sample individuals are mapped to their latent cliques in a  reliable manner.


\begin{table}
\small
\centering
	\begin{tabular}{l  l  cc  cc}
	\toprule
		\multicolumn{5}{l}{\large \textbf{BART}}\\
			\midrule
   &&\multicolumn{2}{c}{\textbf{Training cases}}&\multicolumn{2}{c}{\textbf{Test cases}}\\
			\midrule
   &&B-MSC &Full  &B-MSC &Full \\
   \midrule
   Study 1 &Group 1 &\textbf{38.0} (1.1)  &20.6 (1.0)  &\textbf{37.1} (1.2)  &17.9 (1.1)\\
	&Group 2 &\textbf{52.8} (1.3)  &28.7 (1.1)  &\textbf{51.0} (1.3)  &22.3 (1.3)\\
&Group 3 &\textbf{53.0} (1.3)  &30.4 (1.1)  &\textbf{51.1} (1.3)  &23.2 (1.3)\\
&Group 4 &\textbf{52.5} (1.2)  &29.5 (1.1)  &\textbf{49.8} (1.4)  &22.2 (1.3)\\
Study 2&Group 1  &\textbf{52.8} (1.3)  &31.0 (1.1)  &\textbf{50.2} (1.4)  &24.0 (1.2)\\
&Group 2 &\textbf{52.5} (1.4)  &29.6 (1.1)  &\textbf{49.9} (1.4)  &23.7 (1.3)\\
&Group 3 &\textbf{50.4} (1.3)  &31.5 (1.1)  &\textbf{48.2} (1.4)  &25.4 (1.2) \\
&Group 4 &\textbf{52.4} (1.3)  &29.5 (1.1)  &\textbf{50.5} (1.4)  &23.5 (1.2)\\
\midrule
\multicolumn{5}{l}{\large\textbf{Random forests}}\\
			\midrule
   &&\multicolumn{2}{c}{\textbf{Training cases}}&\multicolumn{2}{c}{\textbf{Test cases}}\\
			\midrule
   &&B-MSC &Full  &B-MSC &Full \\
   \midrule
   Study 1&Group 1 &\textbf{23.3} (0.8)  &19.0 (1.1)  &24.1 (0.9)  &22.5 (1.3)\\
	& Group 2 &\textbf{37.3} (1.1)  &24.8 (1.3)  &\textbf{39.0} (1.2)  &27.2 (1.4)\\
& Group 3 &\textbf{38.9} (1.1)  &29.2 (1.2)  &\textbf{38.6} (1.2)  &32.1 (1.4)\\
& Group 4 &\textbf{37.5} (1.1)  &27.2 (1.3)  &\textbf{38.1} (1.2)  &29.0 (1.5)\\
Study 2 & Group 1 &\textbf{37.1} (1.1)  &27.0 (1.3)  &\textbf{36.7} (1.2)  &30.4 (1.4) \\
& Group 2 &\textbf{37.5} (1.2)  &26.0 (1.3)  &\textbf{37.8} (1.3)  &29.5 (1.4)\\
& Group 3 &\textbf{36.4} (1.1)  &28.9 (1.3)  &\textbf{37.1} (1.2)  &32.1 (1.4)\\
& Group 4 &\textbf{37.9} (1.1)  &25.7 (1.3)  &\textbf{38.8} (1.3)  &29.1 (1.4)\\
		\bottomrule
	\end{tabular}
		\caption{In the \textit{low association} simulation scenario,  accuracy  of inferred o-MPS  utilizing the smaller motif submatrices of the proposed B-MSC method as the covariates compared to the high-dimensional set of covariates (``Full").  
  For \textbf{BART} and \textbf{random forest} estimation procedures (row block) and study-group combination (row), the displayed numbers are the  percentage correlations between the true and estimated o-MPS of the 360 training and 90 test cases (column blocks),   averaged over 500 artificial datasets. Shown in parentheses are the estimated standard errors. Separately for the training samples and test samples of each row, a covariate set (B-MSC or Full) with a significantly higher correlation is highlighted in bold.}\label{table:simulation correlation-1}
\end{table}


\begin{table}
\small
\centering
	\begin{tabular}{l  l  cc  cc}
	\toprule
		\multicolumn{5}{l}{\large \textbf{Ridge regression}}\\
			\midrule
   &&\multicolumn{2}{c}{\textbf{Training cases}}&\multicolumn{2}{c}{\textbf{Test cases}}\\
			\midrule
   &&B-MSC &Full  &B-MSC &Full \\
   \midrule
   Study 1 &Group 1 &\textbf{47.4} (1.7)  &24.4 (1.6)  &\textbf{47.3} (1.7)  &22.4 (1.7)\\
	&Group 2 &\textbf{55.9} (1.6)  &30.8 (1.5)  &\textbf{54.3} (1.6)  &27.1 (1.6)\\
&Group 3  &\textbf{57.8} (1.5)  &33.5 (1.5)  &\textbf{56.6} (1.5)  &29.0 (1.6) \\
&Group 4 &\textbf{56.4} (1.5)  &33.7 (1.4)  &\textbf{53.7} (1.7)  &28.5 (1.6)\\
Study 2&Group 1   &\textbf{56.4} (1.6)  &34.7 (1.5)  &\textbf{54.3} (1.8)  &31.0 (1.6) \\
&Group 2 &\textbf{55.5} (1.6)  &32.1 (1.5)  &\textbf{54.6} (1.7)  &28.4 (1.7)\\
&Group 3 &\textbf{55.2} (1.5)  &36.0 (1.4)  &\textbf{53.9} (1.6)  &33.0 (1.5)\\
&Group 4 &\textbf{56.4} (1.5)  &33.1 (1.5)  &\textbf{55.1} (1.7)  &30.2 (1.6) \\
\midrule
\multicolumn{5}{l}{\large\textbf{Lasso}}\\
			\midrule
   &&\multicolumn{2}{c}{\textbf{Training cases}}&\multicolumn{2}{c}{\textbf{Test cases}}\\
			\midrule
   &&B-MSC &Full  &B-MSC &Full \\
   \midrule
   Study 1 &Group 1 &\textbf{41.9} (1.9)  &21.2 (1.6)  &\textbf{41.3} (1.9)  &19.2 (1.5)\\
	& Group 2  &\textbf{48.6} (1.8)  &24.6 (1.5)  &\textbf{47.2} (1.8)  &20.2 (1.4)\\
& Group 3 &\textbf{50.5} (1.8)  &26.3 (1.6)  &\textbf{49.4} (1.8)  &22.9 (1.6)\\
& Group 4 &\textbf{50.2} (1.6)  &28.2 (1.4)  &\textbf{48.0} (1.8)  &24.3 (1.4)\\
Study 2 & Group 1 &\textbf{50.3} (1.7)  &27.0 (1.5)  &\textbf{48.4} (1.8)  &23.2 (1.4) \\
& Group 2 &\textbf{50.2} (1.7)  &25.7 (1.5)  &\textbf{49.4} (1.7)  &21.1 (1.5)\\
& Group 3 &\textbf{48.3} (1.8)  &28.9 (1.5)  &\textbf{46.8} (1.9)  &25.2 (1.5)\\
& Group 4  &\textbf{50.4} (1.8)  &25.6 (1.7)  &\textbf{49.2} (1.9)  &21.8 (1.6)\\
		\bottomrule
	\end{tabular}
		\caption{In the \textit{low association} simulation scenario,  accuracy  of inferred o-MPS  utilizing the smaller motif submatrices of the proposed B-MSC method as the covariates compared to the high-dimensional set of covariates (``Full"). 
  For \textbf{ridge regression} and \textbf{lasso} estimation procedures (row block) and study-group combination (row), the displayed numbers are the  percentage correlations between the true and estimated o-MPS of the 360 training and 90 test cases (column blocks),   averaged over 500 artificial datasets. Shown in parentheses are the estimated standard errors. Separately for the training samples and test samples of each row, a covariate set (B-MSC or Full) with a significantly higher correlation is highlighted in bold.}\label{table:simulation correlation-2}
\end{table}


\begin{table}
\small
\centering
	\begin{tabular}{l  l  cc  cc}
	\toprule
		\multicolumn{5}{l}{\large \textbf{Adaptive lasso}}\\
			\midrule
   &&\multicolumn{2}{c}{\textbf{Training cases}}&\multicolumn{2}{c}{\textbf{Test cases}}\\
			\midrule
   &&B-MSC &Full  &B-MSC &Full \\
   \midrule
   Study 1 &Group 1 &\textbf{44.3} (1.8)  &22.2 (1.6)  &\textbf{43.2} (1.9)  &20.4 (1.5)\\
	&Group 2 &\textbf{51.5} (1.7)  &26.0 (1.4)  &\textbf{49.8} (1.7)  &20.9 (1.5)\\
&Group 3  &\textbf{53.0} (1.6)  &29.0 (1.5)  &\textbf{51.6} (1.7)  &25.1 (1.6)\\
&Group 4 &\textbf{51.7} (1.6)  &29.1 (1.5)  &\textbf{49.3} (1.8)  &24.5 (1.5)\\
Study 2 &Group 1   &\textbf{52.1} (1.7)  &28.3 (1.5)  &\textbf{49.7} (1.8)  &24.2 (1.4)\\
&Group 2 &\textbf{52.0} (1.6)  &27.7 (1.5)  &\textbf{50.9} (1.7)  &22.9 (1.4)\\
&Group 3 &\textbf{49.5} (1.7)  &31.7 (1.5)  &\textbf{48.3} (1.8)  &28.3 (1.5)\\
&Group 4 &\textbf{52.0} (1.8)  &28.4 (1.6)  &\textbf{51.1} (1.9)  &24.1 (1.7)\\
\midrule
\multicolumn{5}{l}{\large\textbf{Group lasso}}\\
			\midrule
   &&\multicolumn{2}{c}{\textbf{Training cases}}&\multicolumn{2}{c}{\textbf{Test cases}}\\
			\midrule
   &&B-MSC &Full  &B-MSC &Full \\
   \midrule
   Study 1 &Group 1 &\textbf{44.3} (1.8)  &20.5 (1.5)  &\textbf{43.5} (1.8)  &18.6 (1.5)\\
	& Group 2  &\textbf{51.0} (1.7)  &24.6 (1.4)  &\textbf{49.2} (1.7)  &19.7 (1.4)\\
& Group 3 &\textbf{52.5} (1.7)  &26.7 (1.5)  &\textbf{51.4} (1.7)  &22.5 (1.6)\\
& Group 4 &\textbf{52.3} (1.5)  &27.9 (1.5)  &\textbf{49.9} (1.7)  &23.9 (1.4)\\
Study 2 & Group 1 &\textbf{51.2} (1.7)  &26.9 (1.5)  &\textbf{48.7} (1.8)  &22.8 (1.4)\\
& Group 2 &\textbf{51.1} (1.7)  &26.3 (1.5)  &\textbf{50.2} (1.7)  &21.7 (1.4)\\
& Group 3 &\textbf{50.2} (1.7)  &29.3 (1.5)  &\textbf{48.8} (1.8)  &26.4 (1.5)\\
& Group 4  &\textbf{52.2} (1.7)  &25.6 (1.6)  &\textbf{51.3} (1.8)  &20.9 (1.6)\\
		\bottomrule
	\end{tabular}
		\caption{In the \textit{low association} simulation scenario,  accuracy  of inferred o-MPS  utilizing the smaller motif submatrices of the proposed B-MSC method as the covariates compared to the high-dimensional set of covariates (``Full"). 
  For \textbf{adaptive lasso} and \textbf{group lasso} estimation procedures (row block) and study-group combination (row), the displayed numbers are the  percentage correlations between the true and estimated o-MPS of the 360 training and 90 test cases (column blocks),   averaged over 500 artificial datasets. Shown in parentheses are the estimated standard errors. Separately for the training samples and test samples of each row, a covariate set (B-MSC or Full) with a significantly higher correlation is highlighted in bold.}\label{table:simulation correlation-3}
\end{table}


 More specifically,
 when the entire sample of $\tilde{N}$ subjects is used to estimate the unknown B-MSC model parameters, denote the subject-specific latent cliques by $\tilde{r}_i^{[1]}$ and $\tilde{r}_i^{[2]}$, and the motif vectors by $\bigl(\tilde{\boldsymbol{\phi}}^{[1]}_{\tilde{r}_i^{[1]}}, \tilde{\boldsymbol{\phi}}^{[2]}_{\tilde{r}_i^{[2]}}\bigr)$ of length $\tilde{q}_r^{[1]}+\tilde{q}_r^{[2]}$. For the continuous covariates $\mathbf{X}_r^{[1]}$, the parity $\Delta_r^{[1]}$ is defined as the correlation between the motif  element pairs, $\bigl(\tilde{\phi}^{[1]}_{\tilde{r}_i^{[1]}\tilde{c}_j^{[1]}},\,\hat{\phi}^{[1]}_{\hat{r}_i^{[1]}\hat{c}_j^{[1]}}\bigr)$,  over all $j=1,\ldots,p_1$, and  test cases $i$. Notice that the first term depends on latent clique $\tilde{r}_i^{[1]}$ whereas the second term depends on latent clique $\hat{r}_i^{[1]}$.
For the binary covariates $\mathbf{X}_r^{[2]}$, the parity $\Delta_r^{[2]}$ is the proportion of matches between the  motif elements, i.e., the average of $\mathcal{I}\bigl(\tilde{\phi}^{[1]}_{\tilde{r}_i^{[1]}\tilde{c}_j^{[1]}}=\,\hat{\phi}^{[1]}_{\hat{r}_i^{[1]}\hat{c}_j^{[1]}}\bigr)$ over all $j=1,\ldots,p_2$, and test cases $i$. Averaging over the 500 datasets in the low association simulation scenario ($\mu=10$), the average  clique parity $\Delta_r^{[t]}$\% of
test cases was 64.79\% with an estimated standard error of 0.1\% for continuous covariates ($t = 1$), and 97.00\% with an estimated standard error of 0.03\% for binary covariates ($t = 2$). Irrespective of the covariate type, the test case latent clique characteristics when  $\tilde{N}_{\text{train}}$ samples were used to fit the B-MSC model, were   similar to the latent clique characteristics when all $\tilde{N}$ samples were used to train the B-MSC model.
For perspective, in the case of binary covariates, a parity   exceeding 96\% corresponds to fewer than 1,883  mismatches among the 47,070  bits of the test subjects'  motifs, demonstrating the high reliability of inferring the latent cliques. The results were  similar in the high association simulation scenario.

For the $r$th dataset, 
the study-group memberships and estimated motif submatrices of the B-MSC method, estimated using only   the $\tilde{N}_{\text{train}}$ subjects, were used  to     estimate o-MPS using different regression techniques. The  correlation between the  estimate $\hat{\delta}_{szr}(\mathbf{x}_{ir})$ and  true o-MPS,  $\delta_{szr}(\mathbf{x}_{ir})$, $i=1,\ldots,\tilde{N}$, was computed for each $(s,z)$ combination. For the low association scenario ($\mu=10$), the average  correlations over the $R=500$ datasets are reported   in the columns  labeled ``B-MSC'' in Tables \ref{table:simulation correlation-1}-\ref{table:simulation correlation-3}.  For comparison,   the columns  labeled ``Full''  
 display the corresponding  numbers when all $p$  (fully observed) covariates are used as o-MPS predictors. 
Although the multinomial logistic regression model implemented in the {\tt nnet} package was easily able to accommodate the  smaller motif submatrices of B-MSC, the full covariate matrix of dimension $\tilde{N}_{\text{train}}\times$  $p$ was too large to be fit   on a University of Florida HiPerGator2 supercomputer with  Intel E5-2698v3 processors and 4 GB of RAM per core. 
 However, the BART and random forests regression models  analyzed the full set of covariates. In Tables~\ref{table:simulation correlation-1}-\ref{table:simulation correlation-3}, we find that, for most study-group combinations,  and irrespective of the  regression technique, the lower dimensional predictors provided by B-MSC yielded significantly better o-MPS estimates for both training  and test samples, as evidenced by the significantly higher correlations marked in boldface.  In the high association scenario ($\mu=15$), the results were  more strongly in favor of B-MSC;  in Tables~2-4 of the Appendix,  B-MSC    almost uniformly and significantly outperformed the full set of covariates.
 
These results demonstrate the success of   B-MSC  in providing a  lower dimensional representation of  $\mathbf{X}$ with  minimal loss of  information, correctly inferring the  clique memberships of  test cases, and accurately estimating the o-MPS for subsequent covariate-balanced analyses.  This instills confidence in employing the B-MSC technique for analyzing the TCGA data.



\section{Integration and analysis of  breast cancer datasets} \label{S:data analyses}

Integrating the TCGA breast cancer patients from the $J=7$ medical centers,  we aimed to compare the overall survival (OS) of $K=2$ breast cancer subtypes, namely IDC and ILC, among these patients, with the intention of providing insights that can inform the breast cancer population in the United States.
 An evident challenge is the disparity between the percentages of IDC  and ILC patients in the U.S., where they stand at 88.9\% and 11.1\%, respectively \citep{GDC}, which differ much from those in the TCGA studies.
For example, the IGC study comprised only 28.9\% IDC patients. An effective method for integrative analysis  must  account for these discrepancies  while adjusting for  patient attributes such as    clinical,  demographic, and high dimensional biomarkers. However,  integrative combined (IC) weights \citep{guha2023causal}, like the meta-analytical  extensions of most  existing weighting methods, assume a hypothetical pseudo-population with equally prevalent subtypes, i.e.,  50\% IDC and ILC patients, and so does not resemble important aspects of the US patient population. 
 By contrast, the  FLEXOR weights \citep{guha2023causal} guarantee relative weights of  88.9\% and 11.1\%,  in conformity with the subtype prevalence.


\begin{table}[t]
\caption{B-MSC  clusters, along with their allocated covariates, that are predictive of  observed population multiple propensity score (o-MPS) in  the TCGA breast cancer datasets. See the text for a detailed discussion.}\label{T:dataanalysis}
	\begin{tabular*}{\columnwidth}{@{\extracolsep{\fill}}ll@{\extracolsep{\fill}}}
	\toprule
	\textbf{Cluster}   & \textbf{Covariate(s)} \\
	\midrule
	1    & LRRC31, DNMBP-AS1, PYDC1, NPGPR    \\
2 & CCNB3 \\
3     & RPL29P2, RPS26P11  \\
4    & ABCC6P1, SPINK8 \\
5    & WFIKKN2 \\
6    & OR2T8, OR2W8P \\
7    & GDF3, CST2 \\
8    & C1QTNF9B, HBA1 \\
9    & LOC100130264, KAAG1 \\
10    & OR2L13, TAC2, CLDN19, GRIA4, PROL1,\\
    & \hspace{10 pt}  SLC7A3, SLC22A11, RNF186, CDH22 \\
11 & P2RY11, KCNG2, Year of initial diagnosis \\
12 & ERVFRD-1, GATA1, CMA1, GCSAML,   \\
 & \hspace{10 pt} SIGLEC6, SLC8A3, LOC154544, CTSG,  \\
 & \hspace{10 pt} SIGLEC17P, RGS13 \\
13 & C14ORF178, ENPP7, HCG4B, PLA2G1B, \\
 &  RNF113B \\
\hline
14 & KCNB1, SLC9A8, SPATA2, LINC00651,   \\
 &  \hspace{10 pt} UBE2V1, CEBPB, PTPN1, FAM65C,    \\
 & \hspace{10 pt}  MOCS3, PTGIS, B4GALT5, RNF114,  \\
  & \hspace{10 pt}  snoU13, SNAI1, TMEM189, MIR645, \\
   & \hspace{10 pt}  PARD6B, BCAS4, ADNP, DPM1, 4\\
   & \hspace{10 pt}  KCNG1, CBLN4\\
15 & PR+  \\
	\bottomrule
	\end{tabular*}
\end{table}

We analyzed the TCGA breast cancer datasets containing $N=450$ patients, covariate matrix $\mathbf{X}^{[1]}$  with $p_1=757$ continuous  mRNA biomarker   and   clinicopathological measurements,  and  covariate matrix $\mathbf{X}^{[2]}$  with  $p_2=522$ binary CNA biomarker and other covariates, so that $p=1,279$ covariates. For covariate-balanced inferences, the first task was o-MPS estimation.  To implement the proposed B-MSC methodology,  we  generated MCMC samples from the proposed model using the the Appendix algorithm. Following a burn-in period of 10,000 MCMC samples, 50,000 additional  samples were stored. Posterior convergence was confirmed by  trace plots.  We applied the previously described inference strategy to find Bayes estimates    of the   motif submatrices $\mathbf{\Phi}^{[1]}$ and $\mathbf{\Phi}^{[2]}$ with only $\hat{q}_c^{[1]}=184$ and $\hat{q}_c^{[1]}=43$ columns, respectively, corresponding to the estimated clusters. We  applied  the group lasso technique  to
regress $(S_i,Z_i)$ on the lower-dimensional motifs  of length $\hat{q}_c^{[1]} + \hat{q}_c^{[2]}=227$ and so obtain the o-MPS 
 estimates. 
  Table \ref{T:dataanalysis} displays the covariates belonging to the  B-MSC clusters predictive of o-MPS, which may be very different from the predictors of health outcomes such as cancer survival; only 15 of these 227  clusters were predictive of o-MPS. Thirteen predictive clusters corresponded to continuous covariates   predominantly comprising mRNA biomarkers. The only exception was the  year of initial  diagnosis, which (after standardization) exhibited   across-patient patterns similar to the genes P2RY11 and KCNG2. The remaining  clusters consisted of binary covariates;  cluster 14 contained 22 CNA biomarkers and  cluster 15 represented  positive progesterone receptor  status. 

  The biological relevance of the  clusters in Table \ref{T:dataanalysis}  is attested  by the  medical literature. As a  byproduct of the methodology, regressing only the patient disease subtypes on the motifs gives the B-MSC  clusters that differentiate the disease subtypes.  Since out-of-bag prediction, variable selection, and differential analysis are not relevant to o-MPS estimation, the proposed methodology does not focus on identifying   differential covariates. Nevertheless, from the second regression  analysis, we found that group lasso discovered  the same predictor clusters as Table~\ref{T:dataanalysis}. Comparing with the medical literature, we find each predictor cluster consists of known differential biomarkers of IDC and ILC.  Specifically, the biological role of the genes LRRC31, DNMBP, PYDC1, CCNB3, RPL29P2, RPS26P11, SPINK8, WFIKKN2, OR2T8, GDF3, CST2, C1QTNF9B, HBA1, KAAG1, OR2L13, P2RY11, ERVFRD-1, C14ORF178, and snoU13  has been noted  by  scientific investigations listed in  the Appendix. \cite{richer2002differential} show that PR status has an important  biological function in differentiating the disease progression of IDC and ILC.

 Applying the techniques introduced in \cite{guha2023causal}, we obtained the IC, IGO, and FLEXOR weights for the $N=450$ patients in  the seven  TCGA breast cancer datasets. 
 The percent ESS of the IC  pseudo-population was 42.2\% or  189.7 patients. The percent ESS of the IGO  pseudo-population was comparable: 42.1\% or  189.3 patients. By contrast, the FLEXOR pseudo-population had a  higher percent ESS of  81.1\% or  365.0 patients.  
  The optimal amounts of aggregated information   from the seven datasets, listed in Table~1 of the Appendix, were estimated as 22\%,  4\%,  9\%, 13\%, 26\%, 20\%, and 7\%, respectively. By contrast, all the study weights  are inflexibly set to $\frac{100}{7}$\%  in the IC and IGO pseudo-populations and may be suboptimal for integrative analyses.

 \begin{table}
\caption{
 Covariate-specific absolute standardized biases in the TCGA breast cancer studies for three pseudo-populations. }
 \label{tab:10}
 \begin{adjustwidth}{-65 pt}{}
\begin{center}
\begin{tabular}{llll}
\\
  \toprule
  &\textbf{FLEXOR} &\textbf{IGO} &\textbf{IC} 
		\\
\midrule
Age at diagnosis & 3.6 & 5.2 & 4.0 \\
Cancer in nearby lymph nodes & 2.8 & 3.8 & 3.1 \\
Percentage genome altered & 4.0 & 2.9 & 3.8 \\
Year of diagnosis & 5.8 & 5.3 & 5.6 \\
Menopause status 1 or 2 & 5.1 & 3.8 & 5.1 \\
Cancer stage 1 or 2 & 2.2 & 2.6 & 2.1 \\
Positive ER status & 3.7 & 3.2 & 3.6 \\
Positive PR status & 5.2 & 4.1 & 5.3 \\
LRRC31 & 2.9 & 3.1 & 3.0 \\
CCNB3 & 4.8 & 3.9 & 4.3 \\
RPL29P2 & 3.9 & 3.1 & 4.1 \\
HBA1 & 4.6 & 3.2 & 4.3 \\
	 \bottomrule
\end{tabular}
\end{center}
\end{adjustwidth}
\end{table}

\begin{figure}
\begin{center}
\includegraphics[scale=0.3]{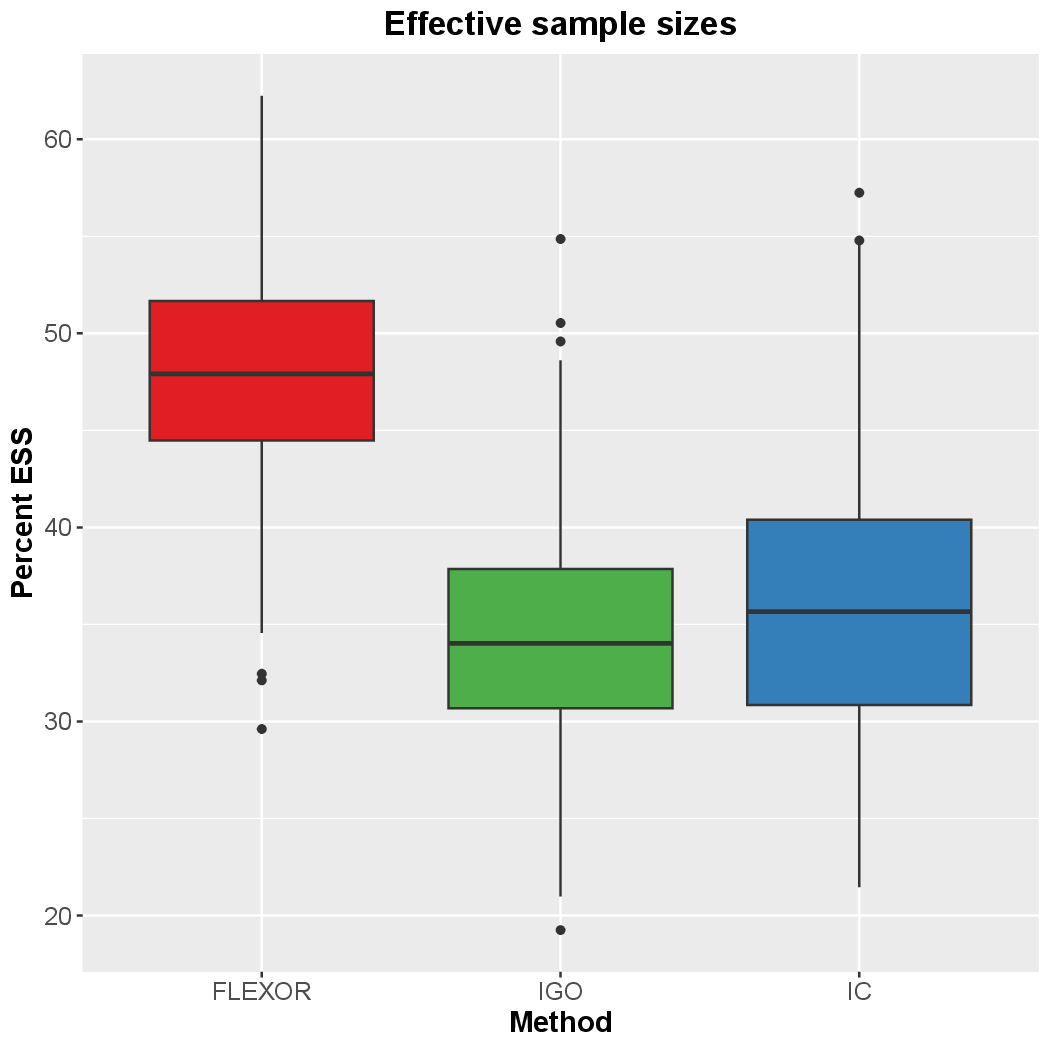}
\includegraphics[scale=0.3]{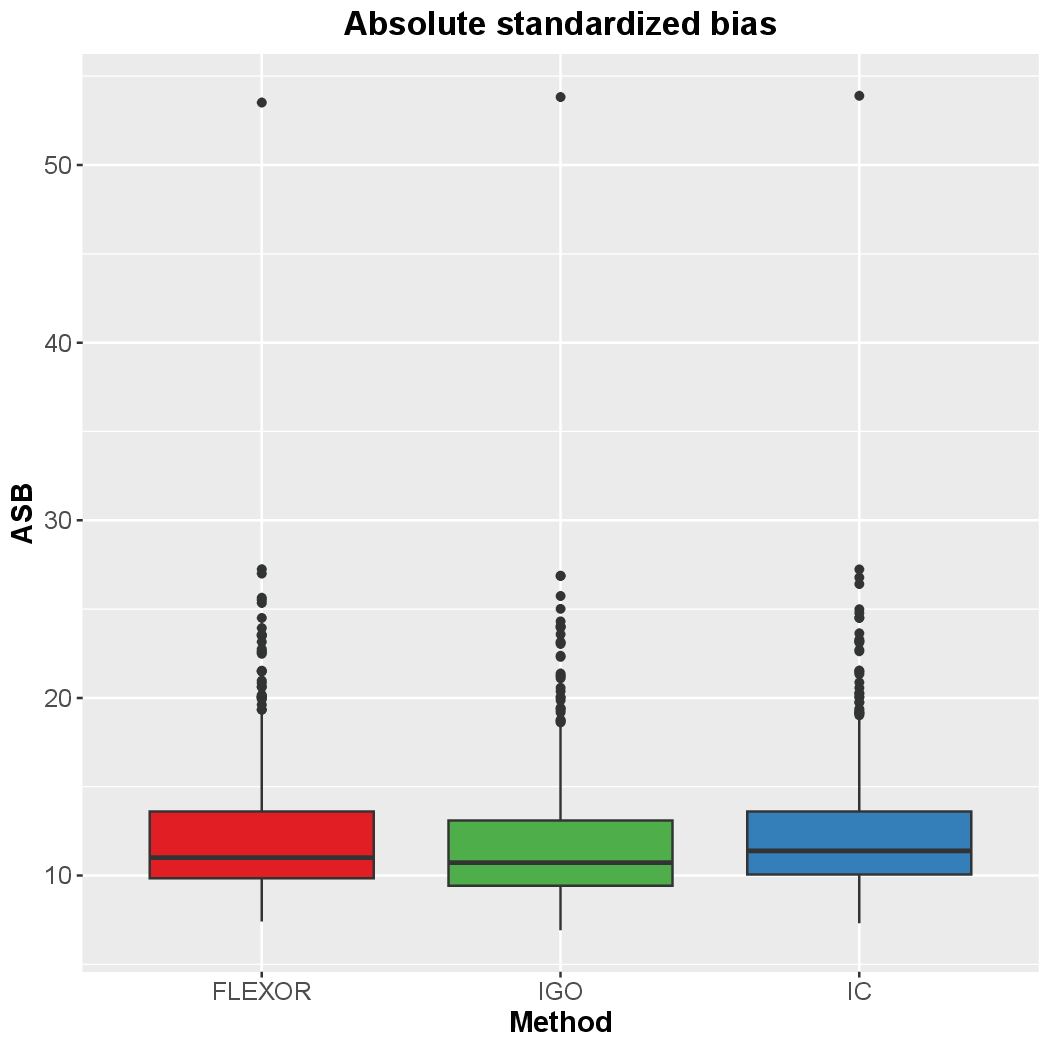}
\caption{For 500 bootstrap samples of 450 patients each drawn from the TCGA breast cancer datasets,  side-by-side boxplots of the  percent ESS (left panel) and   absolute standardized bias (right panel) for the IC, IGO, and FLEXOR pseudo-populations.}
\label{F:TCGA_boxplots}
\end{center}
\end{figure}

 We analyzed 500 bootstrap samples of 450 patients  drawn with replacement from the TCGA breast cancer database.   Side-by-side boxplots of the  percent ESS of the three weighting methods are displayed in the left panel of Figure \ref{F:TCGA_boxplots}. For the bootstrap samples, we find that the FLEXOR pseudo-population had a significantly larger ESS than the other pseudo-populations, indicating this weighting method may be more precise for  wide-ranging pseudo-population estimands such as percentiles and comparative  group survival features such as differences of the mean or median survival times. Boxplots of the    absolute standardized bias \citep{Li_etal_2018} are shown in the right panel of Figure \ref{F:TCGA_boxplots}. Smaller ASB are indicative of better covariate balance. We find that  the weighting methods are  equally effective in mitigating imbalances in the large number of covariates.  We observe a comparable outcome in Table~\ref{tab:10}, showing the ASBs of certain covariates from the  TCGA database.
Finally,  estimates of the survival functions of the    $K=2$ disease subtypes (IDC and ILC) were meta-analyzed using different weighting methods as follows. 
 BKME (\ref{BKME}) was  evaluated using these quantities.
Uncertainty estimation was based on $B=500$ bootstrap samples.


\begin{figure}
\includegraphics[scale=0.3]{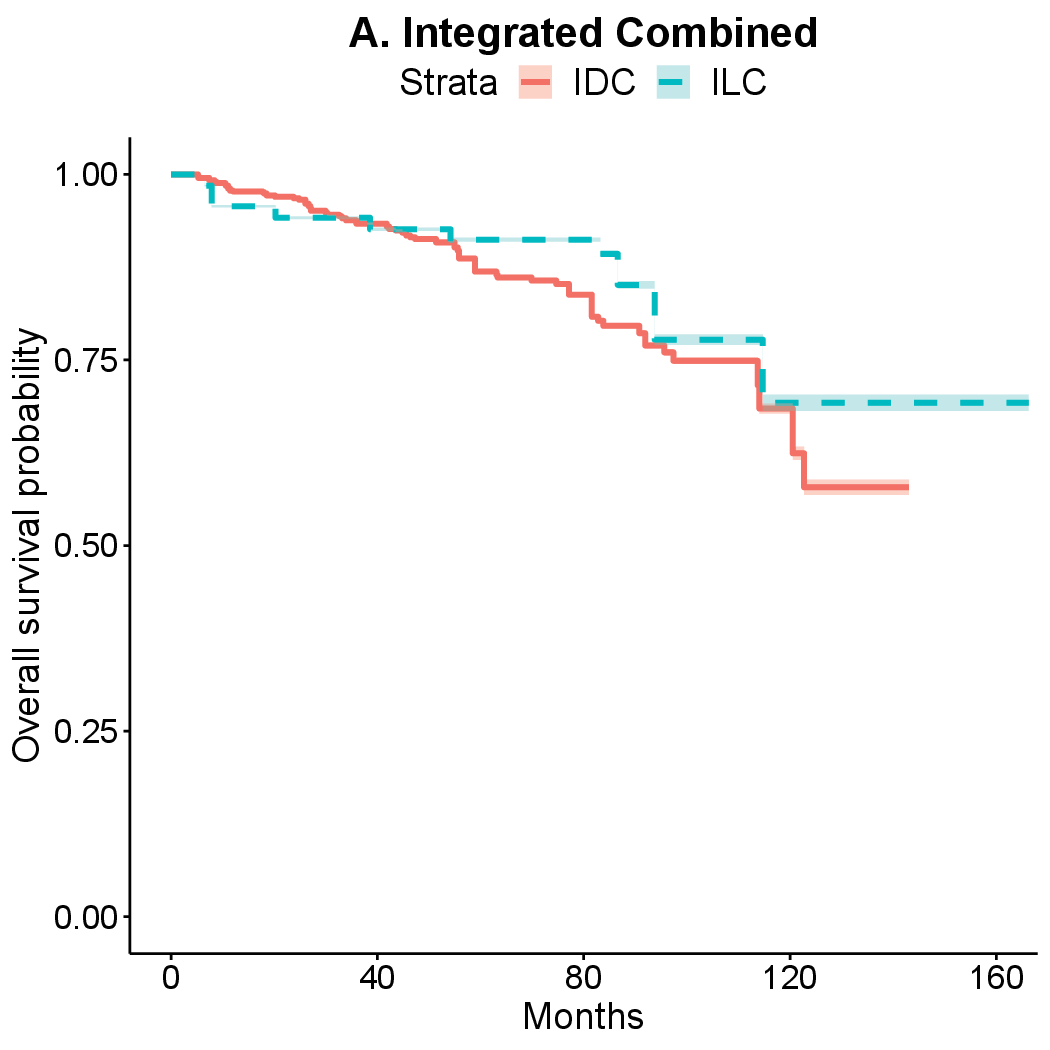}
\includegraphics[scale=0.3]{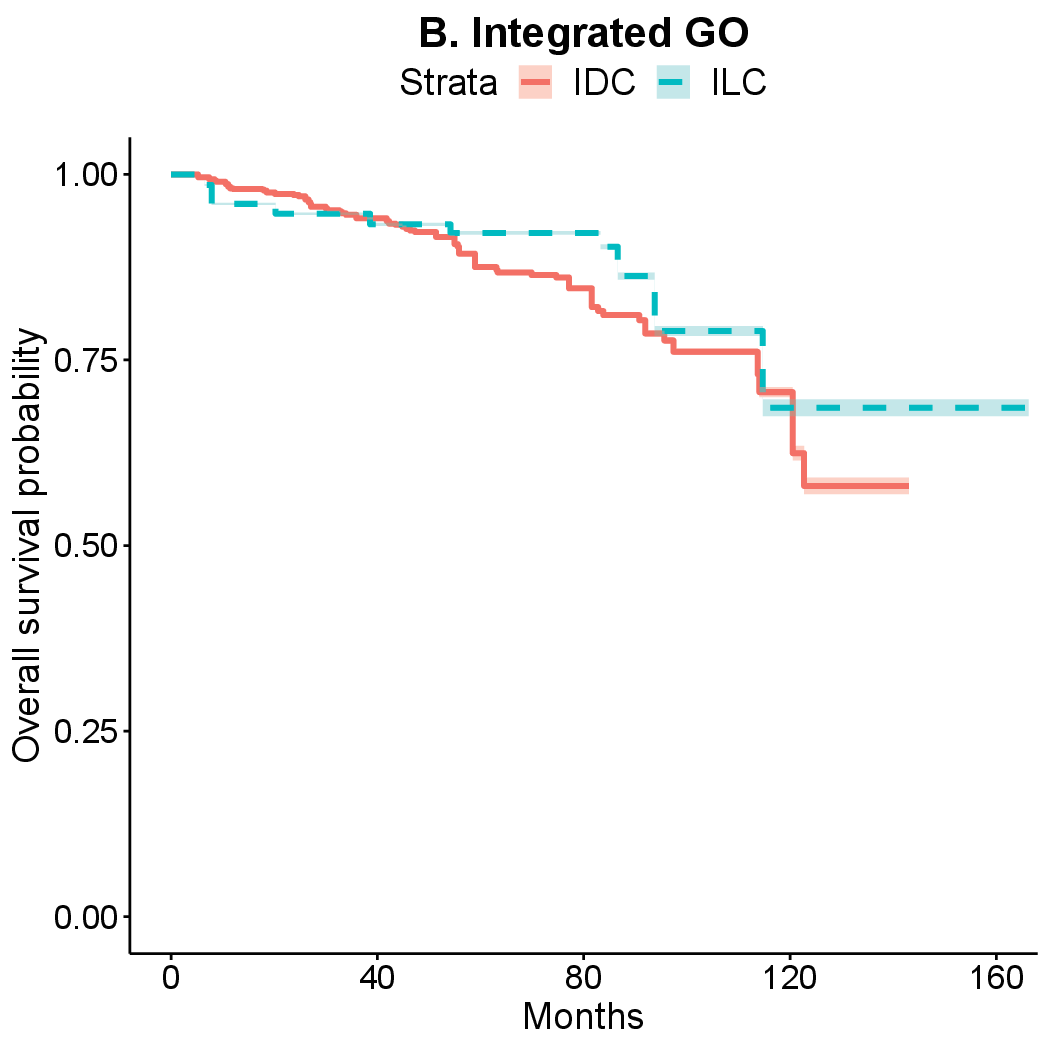}
\includegraphics[scale=0.3]{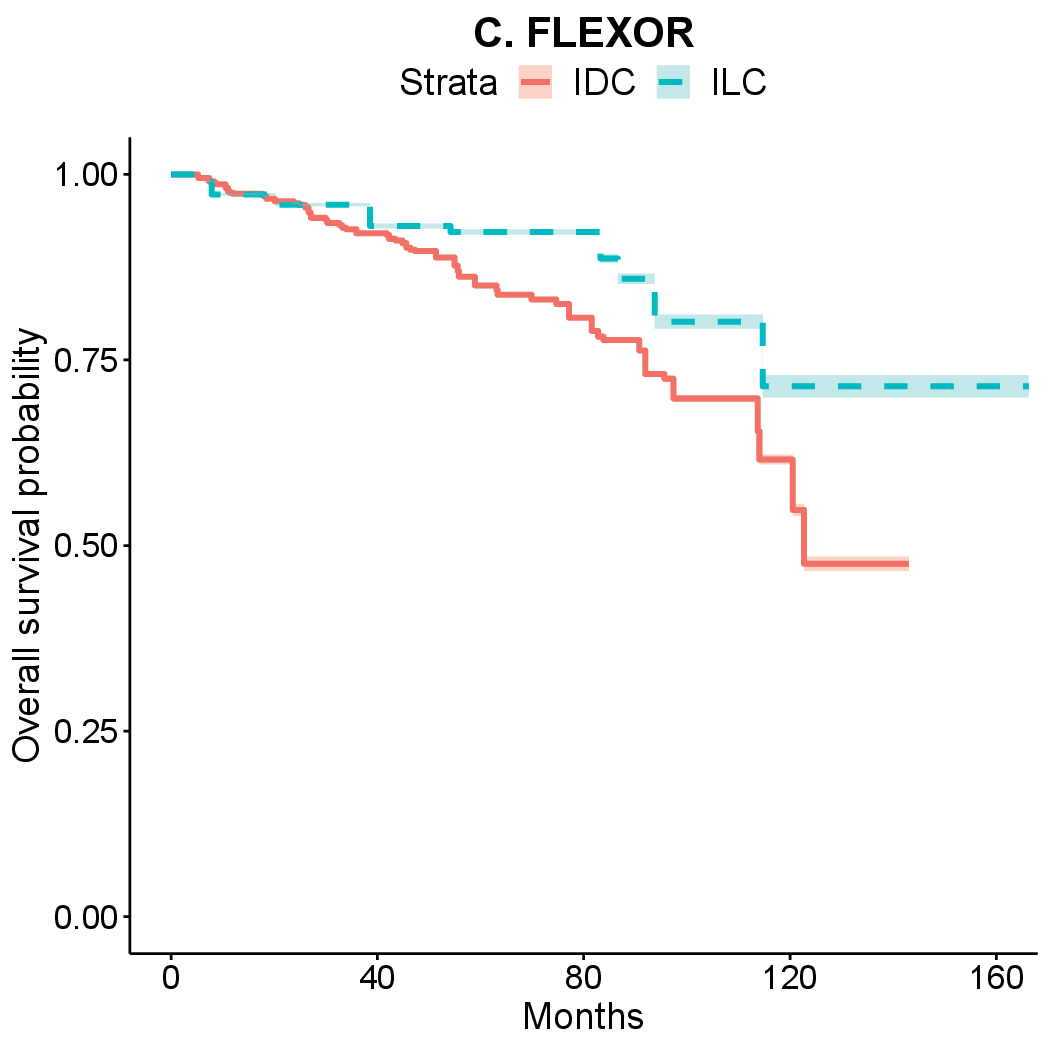}
\caption{For the TCGA breast cancer patients, estimated overall survival  (OS) curves  for disease subtypes IDC (red, solid lines) and ILC (cyan, dashed lines)  with integrative combined (IC), integrative generalized overlap (IGO), and   FLEXOR weights. The standard error bands were estimated from 500 bootstrap samples and are shown in pink and light cyan respectively for IDC and ILC.}\label{F:compare}
\end{figure}


 Adjusting for  stages and grades, Figures \ref{F:compare}A--\ref{F:compare}C provide a comparative analysis between IDC and ILC patients using three different weighting methods: IC, IGO, and FLEXOR. Our findings reveal that the IC and IGO weighting methods yield similar conclusions regarding survival times, suggesting inconclusively that IDC generally exhibits a poorer prognosis than ILC—although this trend may not hold true for lower Overall Survival (OS) rates. Notably, both subtypes show survival probabilities exceeding 50\%.
In contrast, the FLEXOR weighting method indicates significantly worse health outcomes for IDC compared to the other methods. For instance, the 10th percentile of OS for IC, IGO, and FLEXOR weights were 55.0, 55.8, and 46.4 months, respectively—highlighting statistically significant differences when accounting for standard errors. More importantly, FLEXOR consistently suggests that IDC has uniformly worse outcomes than ILC, regardless of OS, with a median IDC survival time of 122.7 months (SE: 0.1 months).
The difference in results may be attributed to the considerably lower effective sample size (ESS) and  somewhat unrealistic  assumptions (e.g., assuming equally prevalent disease subtypes) of the IC and IGO pseudo-populations. This raises questions about the validity of IC and IGO inferences for these datasets, emphasizing the need for careful consideration of methodological assumptions and their potential impact on study outcomes.  Conversely, identifying IDC as consistently associated with poorer outcomes than ILC  (Figure \ref{F:compare}C)  could pave the way for more precise and targeted therapeutic interventions, potentially benefiting both patient groups.

\section{Conclusion} \label{S:discussion}

Propensity scores  play a pivotal role, serving as the foundation for  weighting or matching methods for covariate-balanced analyses. A primary barrier for integrating retrospective  cohorts is  covariate imbalance across the studies and  groups. 
In the context of integrative analyses encompassing multiple observational studies with diverse and unbalanced groups, the concept of propensity scores was generalized by \cite{guha2023causal} to the observed population multiple propensity score (o-MPS); this is defined as the probability of a study-group combination given the subject's covariates. 
However, the presence of high-dimensional covariates introduces complexity in estimating the o-MPS and, consequently, poses a challenge in integrating observational studies with substantial number of covariates.

We have proposed a novel, hybrid  Bayesian-frequentist technique called B-MSC. Exploiting the dimension-reduction property of non-parametric CRPs, we discover latent lower-dimensional archetypes in the covariates called motifs. Using these motifs as potential regressors, standard regularization  techniques can be employed to accurately and flexibly estimate the  propensity scores. 
Using a computationally efficient MCMC algorithm, we foster an inferential procedure that discovers  motif matrices associated with high-dimensional covariates to accurately estimate the o-MPS. We then apply these techniques to make covariate-balanced weighted inferences  using censored survival outcomes in integrative analyses of high-dimensional TCGA breast cancer studies. 

The  B-MSC methodology is capable of accommodating a blend of retrospective cohorts and RCTs. 
 In  future research, we will   extend this strategy to  
transportability  \citep{westreich2017transportability} and data-fusion \citep{dahabreh2023efficient} problems, which utilize random samples  from the  natural population.


\bibliographystyle{agsm}
\bibliography{AllRefs_mar2022, biological}

\end{document}